\documentclass[aps,pre,reprint,showpacs,superscriptaddress]{revtex4-1}
\usepackage{graphicx}
\usepackage{dcolumn}
\usepackage{bm}
\usepackage{color}
\usepackage{lipsum}
\usepackage{mathrsfs}
\usepackage{lineno}
\usepackage{siunitx}
\usepackage{dcolumn}
\usepackage{amsmath}
\usepackage{hyperref}
\hypersetup{hypertex=true,colorlinks=true,linkcolor=blue,anchorcolor=blue,citecolor=blue,urlcolor=blue}
\usepackage{enumerate}
\usepackage{multirow}
\usepackage{appendix}
\usepackage{physics}
\begin{document}

\title{Stopping power of high-density alpha-particle clusters in warm dense deuterium-tritium fuels}

\author{Z. P. Fu}
\affiliation{Key Laboratory for Laser Plasmas, School of Physics and Astronomy, and Collaborative Innovation Center of IFSA (CICIFSA), Shanghai Jiao Tong University, Shanghai 200240, People’s Republic of China}
\affiliation{Zhiyuan College, Shanghai Jiao Tong University, Shanghai 200240, People’s Republic of China}
\author{Z. W. Zhang}
\affiliation{Key Laboratory for Laser Plasmas, School of Physics and Astronomy, and Collaborative Innovation Center of IFSA (CICIFSA), Shanghai Jiao Tong University, Shanghai 200240, People’s Republic of China}
\affiliation{Zhiyuan College, Shanghai Jiao Tong University, Shanghai 200240, People’s Republic of China}
\author{K. Lin}
\affiliation{Key Laboratory for Laser Plasmas, School of Physics and Astronomy, and Collaborative Innovation Center of IFSA (CICIFSA), Shanghai Jiao Tong University, Shanghai 200240, People’s Republic of China}
\affiliation{Zhiyuan College, Shanghai Jiao Tong University, Shanghai 200240, People’s Republic of China}

\author{D. Wu}
\email{dwu.phys@sjtu.edu.cn}
\affiliation{Key Laboratory for Laser Plasmas, School of Physics and Astronomy, and Collaborative Innovation Center of IFSA (CICIFSA), Shanghai Jiao Tong University, Shanghai 200240, People’s Republic of China}
\affiliation{Zhiyuan College, Shanghai Jiao Tong University, Shanghai 200240, People’s Republic of China}

\author{J. Zhang}
\email{jzhang@iphy.ac.cn}
\affiliation{Key Laboratory for Laser Plasmas, School of Physics and Astronomy, and Collaborative Innovation Center of IFSA (CICIFSA), Shanghai Jiao Tong University, Shanghai 200240, People’s Republic of China}
\affiliation{Zhiyuan College, Shanghai Jiao Tong University, Shanghai 200240, People’s Republic of China}
\affiliation{Institute of Physics, Chinese Academy of Sciences, Beijing 100190, People’s Republic of China}
\date{\today}

\begin{abstract}
The state of burning plasma had been achieved in inertial confinement fusion (ICF), which was regarded as a great milestone for high-gain laser fusion energy. In the burning plasma, alpha particles incident on the cryogenic (warm dense) fuels cannot be simply regarded as single particles, and the new physics brought about by the density effects of alpha particles should be considered. In this work, the collective interaction between them has been considered, namely the effect of the superposition of wake waves. The stopping power of alpha-particle clusters, i.e. the rate of energy loss per unit distance traveled has been calculated using both analytical and simulation approaches. In theory, we obtain the stopping power of alpha clusters in cryogenic (warm dense) fuel by the dielectric function method, which illustrates the importance of the effective interaction between particles. Simulation results using the LAPINS code show that the collective stopping power of the alpha cluster is indeed increased via coherent superposition of excitation fields (the excitation of high-amplitude wake waves). However, the comparison between simulation and theoretical results also illustrates a coherent-decoherent transition of the stopping power of the cluster. The initial conditions with various sizes and densities of the alpha clusters have been considered to verify the condition of decoherence transition. Our work provides a theoretical description of the transport properties of high-density alpha particles in warm dense cryogenic fuel and might give some theoretical guidance for the design of actual fusion processes.
\end{abstract}

\maketitle

\section{Introduction}\label{sec:first}

Recently, burning plasma and ``ignition'' in the laboratory had been successively achieved  in the inertial confinement fusion (ICF) approach \cite{r1,r2,r2-1,r0,r0-1,r0-2}, which were regarded as along-standing milestones due to the promise of clean limitless energy.
To achieve these, the central low-density deuterium-tritium (DT) gas (“hot spot”) of the assembled target is compressed and heated to a real density of more than
0.4 g/cm$^{-2}$ and temperature of more than 5 keV using a higher density DT fuel piston accelerated to high velocities ($\sim400$ km/s) that does mechanical work on the hot spot. Under burning conditions, the fusion reaction in the hot spot is violent, producing a large number of alpha particles. 
Such alpha particles incident on the high-density fuels cannot be simply regarded as intermittent single particles. Relevant density effects induced by bulks of particles (“clusters”) should be taken into account.

\begin{figure}[h]
\includegraphics[scale=0.5]{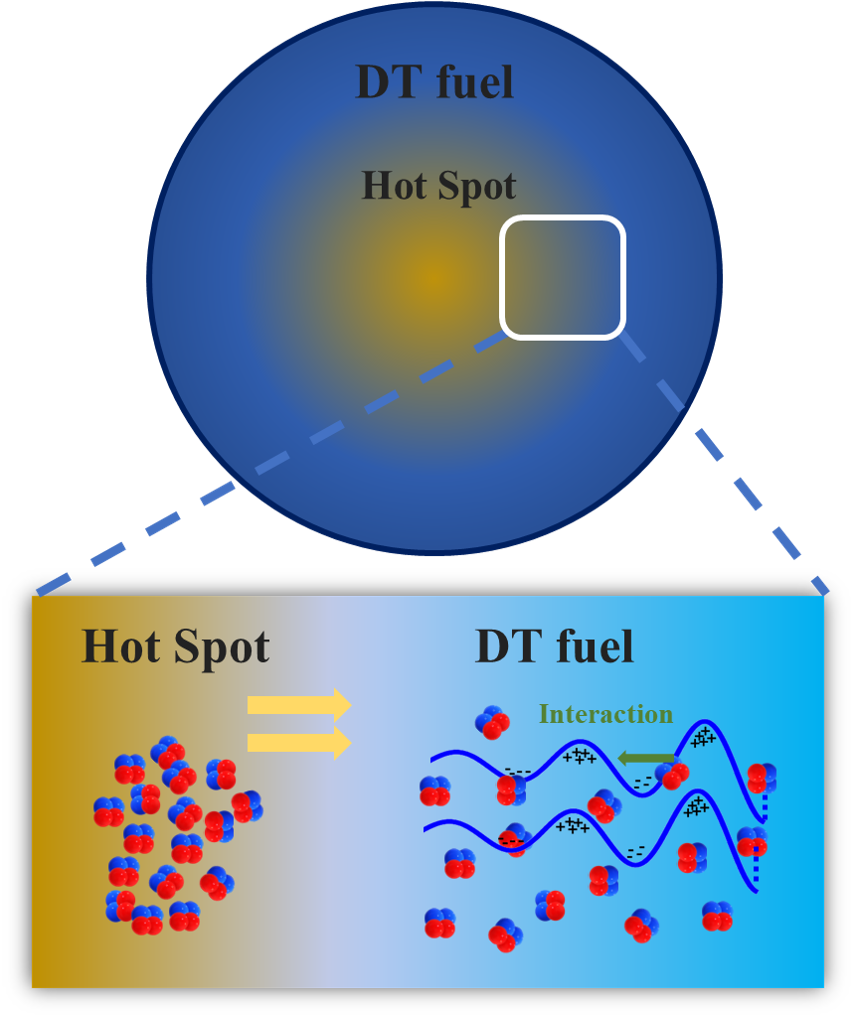}
\caption{\label{fig:figure0} Transport and stopping of alpha clusters in warm dense DT fuel. The blue line represents the electronic field generated by alpha particles. The green arrow indicates the interaction be- tween different alpha particles.}
\end{figure}

\begin{figure}[h]
\includegraphics[scale=0.38]{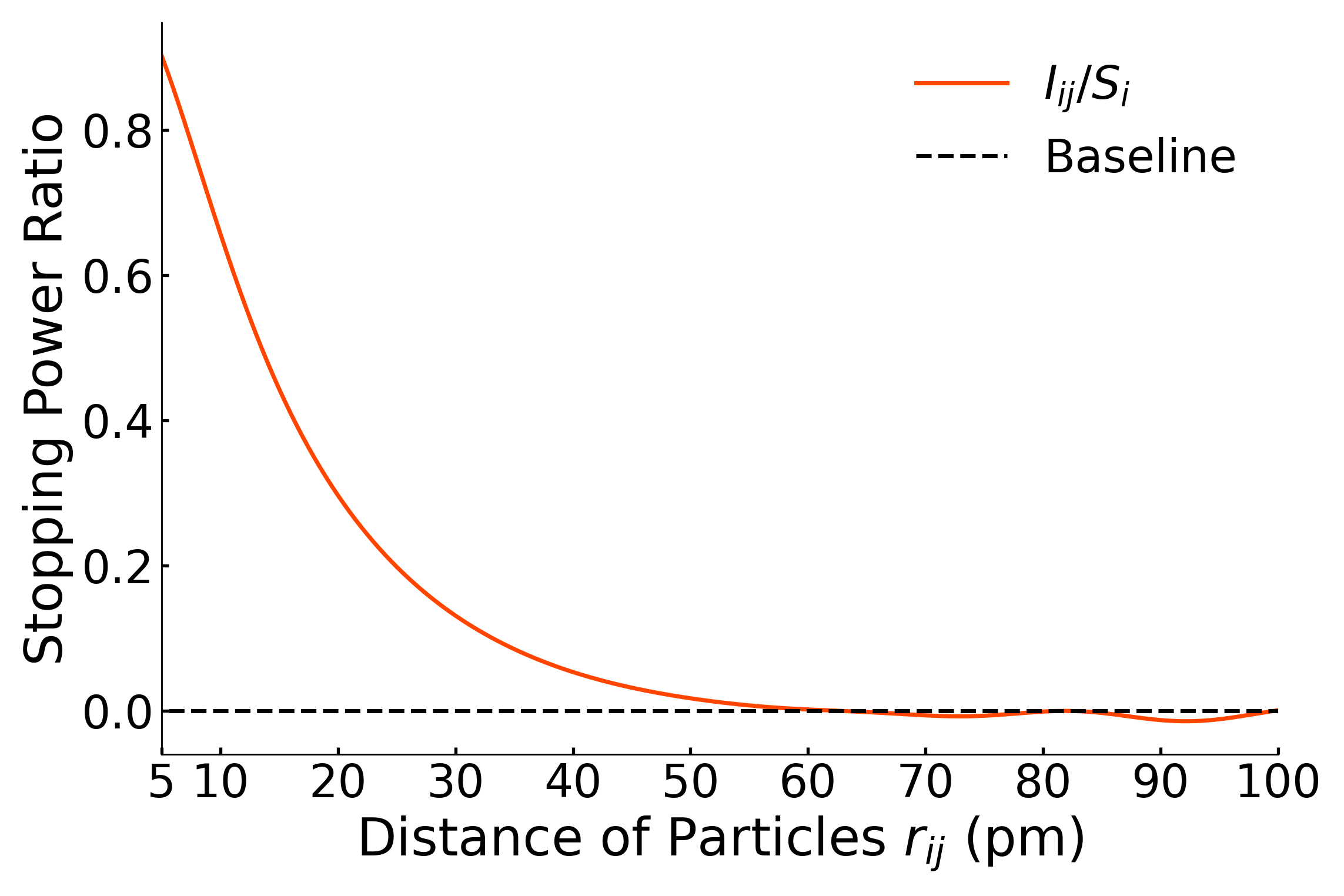}
\caption{\label{fig:figure1} The decaying behaviour of the interference stopping power $I_{ij}$ with the distance between particles $r_{ij}$. The interference stopping power vanishes for $r_{ij}$ beyond $\sim 60~\mathrm{pm}$, leaving behind negligible fluctuations.}
\end{figure}

Stopping power refers to the rate at which a particle loses energy as it passes through plasma, and it is a critical parameter for predicting the transport behavior of particles in plasma. 
Stopping power will affect the range of alpha particles in DT fuel, which in turn affects the propagation of the burning wave and burning efficiency of the fuel \cite{r0}.
Accurate stopping power is therefore a key component for the modeling of ICF.
Studying the energy deposition process of high-density alpha clusters under burning conditions is therefore of great importance for the prediction of burning wave propagation into the high-density DT fuel, which is required for high energy gain in the ICF approach.

Physically, the stopping power of alpha particles in the high-density DT fuel, which usually is at the so-called warm dense matter (WDM) state \cite{wdm1,wdm2,wdm3}, can be divided into three parts–the scattering of nearly-free electrons, the bound electronic stopping power, and the interaction between the alpha particle and ions.  Since the background is dense DT plasmas, bound electrons can be neglected. The background electron in plasma should be considered as partially degenerate Fermi gas \cite{r1,r2}, since the temperature of the cold fuel is lower than the Fermi temperature.

A lot of methods can be used for ab-initio investigation of particle stopping power. For example, Y. H. Ding et al developed time-dependent orbital-free density functional theory (TD-OF-DFT) method \cite{r3}.  Given the high density, the alpha cluster cannot be considered as single particles without interactions. The stopping power of an alpha cluster has also been calculated \cite{r4,r5} using the dielectric method considering the density effects. However, in burning plasma, we should consider our background plasma as the degenerate DT fuel.

In this paper, using the dielectric method, the interaction of alpha particles is studied. The recently developed PIC code LAPINS \cite{r7,r8,r9} is used to verify our theoretical predictions. We found that in WDM when the incident alpha particle was considered a cluster, a larger stopping power would take place due to the superposition of the wake fields generated by different alpha particles in the cluster. In low-density and small-size regime, as the density and the thickness of the cluster increase, this effect becomes stronger and accounts for a significant portion of the stopping power in WDM. In addition, the comparison between simulation and theoretical results also illustrates a decoherence transition of the stopping power for high-density and large-size clusters. 
Our work provides a theoretical description of the transport properties of high-density alpha clusters in WDM and guides the design of actual fusion processes from a theoretical perspective.

The paper is organized as follows. Sec.~\ref{sec:model} introduces the theoretical derivation and simulation setup parameters. In Sec.~\ref{sec:discuss}, based on 1D simulation result of LAPINS, we show the decoherence phenomenon and illustrate the genesis qualitatively. The conclusions are summarized in Sec. ~\ref{sec:sum}.

\section{Model description} \label{sec:model}

In this section, we first provide a theoretical description of the stopping power $S = -\left\langle{dE}/{dx}\right\rangle$ of alpha clusters and then introduce our numerical simulations with the LAPINS code.
\subsection{Theoretical analysis}

\begin{figure*}[htbp]
    \centering
    \includegraphics[scale=0.275]{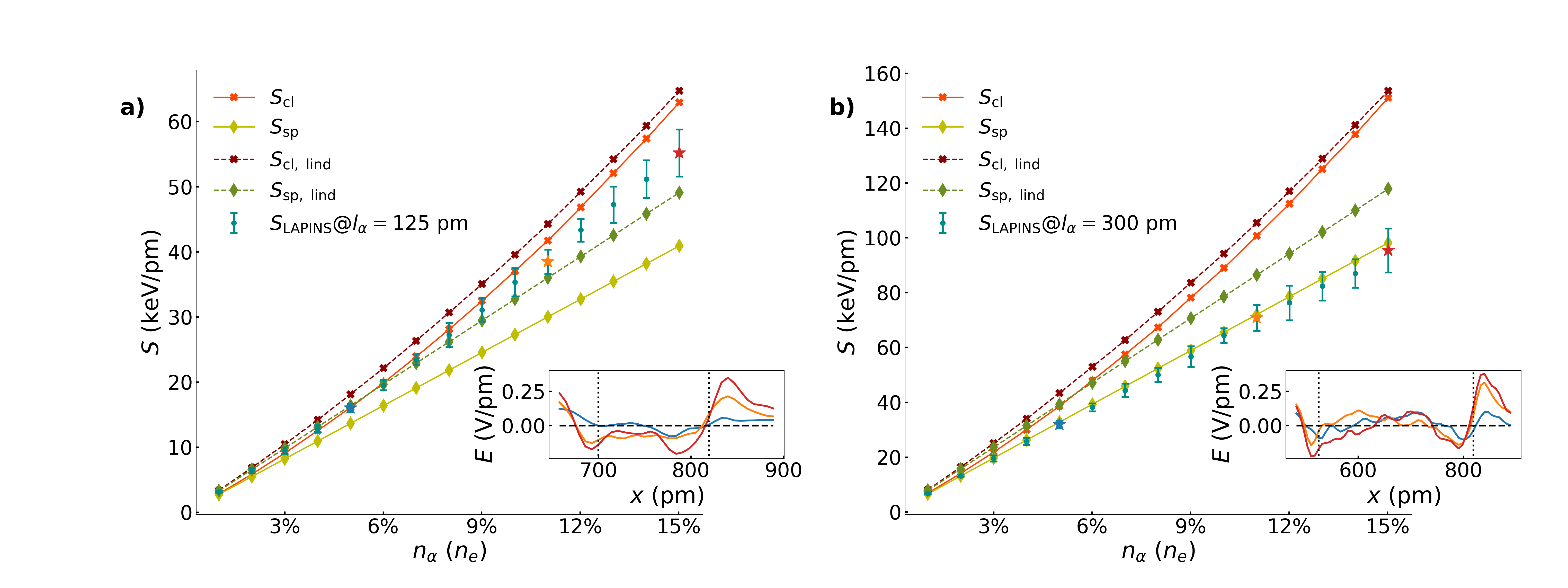}
    \caption{Stopping power of two alpha clusters with different densities, a) \label{fig:125pm} simulation results when $l_\alpha=125~\mathrm{pm}$, b) \label{fig:300pm} simulation results when $l_\alpha=300~\mathrm{pm}$. In the main plots, green dots and error bars: LAPINS simulation result; Red line: interaction term using dielectric method; Yellow line: single particle term using dielectric method. The dash line represents the result using Lindhald dielectric function, while the solid line represents the result of temperature-dependent dielectric function. Error bars come from stopping power at different simulation times, and the confidence interval is 95$\%$. In the insets, simulation results for the electric fields along the longitudinal direction inside the clusters at $t=66.8\times 10^{-3}~\mathrm{fs}$ are shown. Line colors represent the parameters of the corresponding simulation, which can be read from starred points of the same color in the main plots, giving $n=0.05n_e, 0.08n_e, 0.15n_e$ for blue, orange, red lines respectively. And the dash line in the insets represents the location of the cluster.}
    \label{fig:FixedSize}
\end{figure*}

In the implosion process of ICF, by adjusting the laser waveform, a near-isentropic compression can be realized \cite{r0-1,r0-2}, enabling the presence of fuel with high density and low temperature, a crucial condition for the fusion process. Such temperature of the fuel in a fusion process is usually lower than the Fermi temperature of the background electron. Therefore, the background electrons partially degenerate, making it necessary to introduce quantum descriptions. Within the framework of linear response theory, incident alpha particles are regarded as electrostatic perturbation and the quantum properties of electrons can be encoded with the dielectric function \cite{r4,r5}. The stopping power of alpha clusters can be expressed as 
\begin{equation}
		S_{cl}(T,\vb{v}) = \sum_{i=1}^{N}S_i(T,\vb{v})+\sum_{i\ne j}I_{ij}(\vb{r}_{ij},\vb{v})
	\end{equation}
	where
	\begin{equation}
	\begin{aligned}
	    S_i(T,\vb{v})&=\frac{Z^2e^2}{2\pi^2v}\int d^3\vb{q}\frac{\vb{q}\cdot\vb{v}}{q^2}\operatorname{Im}[\frac{-1}{\epsilon(q,\vb{q}\cdot\vb{v},T)}]\\
	    &=\frac{2Z^2e^2}{\pi v^2}\int_{0}^{\infty}\frac{dq}{q}\int_{0}^{qv}d\omega \omega \operatorname{Im}[\frac{-1}{\epsilon(q,\omega,T)}]\\
	\end{aligned}
	\end{equation}
and
	\begin{equation}
	\begin{aligned}
	    I_{ij}(\vb{r}_{ij},T,\vb{v})=&\frac{Z^2e^2}{2\pi^2v}\int d^3\vb{q}\frac{\vb{q}\cdot\vb{v}}{q^2}\\
	    & \times \operatorname{Im}[\frac{-1}{\epsilon(q,\vb{q}\cdot\vb{v},T)}]e^{i\vb{q}\cdot\vb{r}_{ij}}.\\
	\end{aligned}
	\end{equation}
	
Here $S_i$ represents the stopping power of a single incident particle from its interaction with the plasma and $I_{ij}$ corresponds to interference stopping power, which represents the retardation effect of the excitation field of one particle on another particle. Here $v$ represents the velocity of the cluster and $\vb{r}_{ij}$ refers to the distance of two particles in the cluster.
In the formulae, the dielectric function comes from the second quantized Hamiltonian of the Coulomb electron gas with RPA approximation \cite{r15,r16}, which is a classic problem in condensed matter physics. Given that the background plasma temperature in the problem we are studying is at the order of $100~\mathrm{eV}$, we consider the influence of temperature on the dielectric function here \cite{r17,r18,hu,liang1,liang2} instead of using the form of the Lindhard dielectric function at zero temperature,

\begin{equation}
    \epsilon(q,\omega,T) = 1+\frac{3n_ee^2}{2\epsilon_0\varepsilon_fq^2}\chi
\end{equation}
where
\begin{equation}\label{eq1}
    \operatorname{Im}\chi(x,y,z) = \frac{z \pi}{4 x} \ln \left[\frac{1+\mathrm{e}^{-\left(\alpha_{1}^{2}-\gamma\right) / z}}{1+\mathrm{e}^{-\left(\alpha_{2}^{2}-\gamma\right) / z}}\right]
\end{equation}
and
\begin{equation}
    \operatorname{Re}\chi(x,y,z)=\frac{1}{\pi} P \int_{-\infty}^{+\infty} \frac{\operatorname{Im} \chi^{0}\left(x, y^{\prime}, z\right)}{y^{\prime}-y} \mathrm{~d} y^{\prime}.
\end{equation}
In Eq. \ref{eq1}, $x = q/k_f$, $y = \hbar\omega/\varepsilon_f$, $z=k_BT/\varepsilon_f$, $\gamma = \mu/\varepsilon_f$, $\alpha_1 = 1/2((y/x)-x)$ and $\alpha_2 = 1/2((y/x)+x)$, where $k_f$ and $\varepsilon_f$ are Fermi wave vector and Fermi energy, $\mu$ is the chemical potential and $T$ is the temperature of the background plasma.

We work on the scenario that high-density alpha clusters are ejected along the normal direction of the central hot spot spherical surface. Therefore, clusters can be regarded locally as uniformly distributed particles, and the scale in the $x$ and $y$ directions is much larger than the scale in the $z$ direction ($z$ is the normal direction of the surface), which enables the one-dimensional simulation we use. Moreover, because of the uniform density distribution, the random orientation of $\vb{r}_{ij}$ can be averaged out when considering the interference stopping power $I_{ij}$ between pairs of particles, so that $I_{ij}$ depends only on the particle distances.
\begin{equation}
\begin{aligned}
    I_{ij}(r,T,v) =& \frac{2Z^2e^2}{\pi v^2}\int_{0}^{\infty}\frac{dq}{q}\frac{\sin(qr)}{qr}\\
    &\times \int_0^{qv}d\omega \omega \operatorname{Im}[\frac{-1}{\epsilon(q,\omega,T)}]\\
\end{aligned}
\end{equation}

A simple picture illustrating the interference stopping power will be a coherent superposition between the wake fields in the background plasma excited by various single particles, which increases the blocking force despite the local electric field excited by single-particles. In order to better understand this kind of interaction effect, we demonstrate the wake field in the case of coherent superposition as shown in Fig.\ \ref{fig:figure0}\ .

In the estimation of this theoretical model, we randomly place the particles representing the components of the clusters on the site of a grid which has the same size as the whole particle cluster, numerically calculate the dielectric function at the initial temperature of the background plasma, and evaluate the individual stopping power and interference stopping power of each particle pair. Then the stopping power of the total cluster is obtained by adding up the stopping power of each pair. 
The length of the grid and a reasonable truncation length of $I_{ij}$ is determined by the decay length of the $I_{ij}$, shown in Fig.\ \ref{fig:figure1}\ . Based on the result, we take the distance between the grid points as $0.005~\mathrm{nm}$, and truncate $I_{ij}$ when the distance between two particles reaches $0.1~\mathrm{nm}$.

\begin{figure*}[htbp]
    \centering
    \includegraphics[scale=0.275]{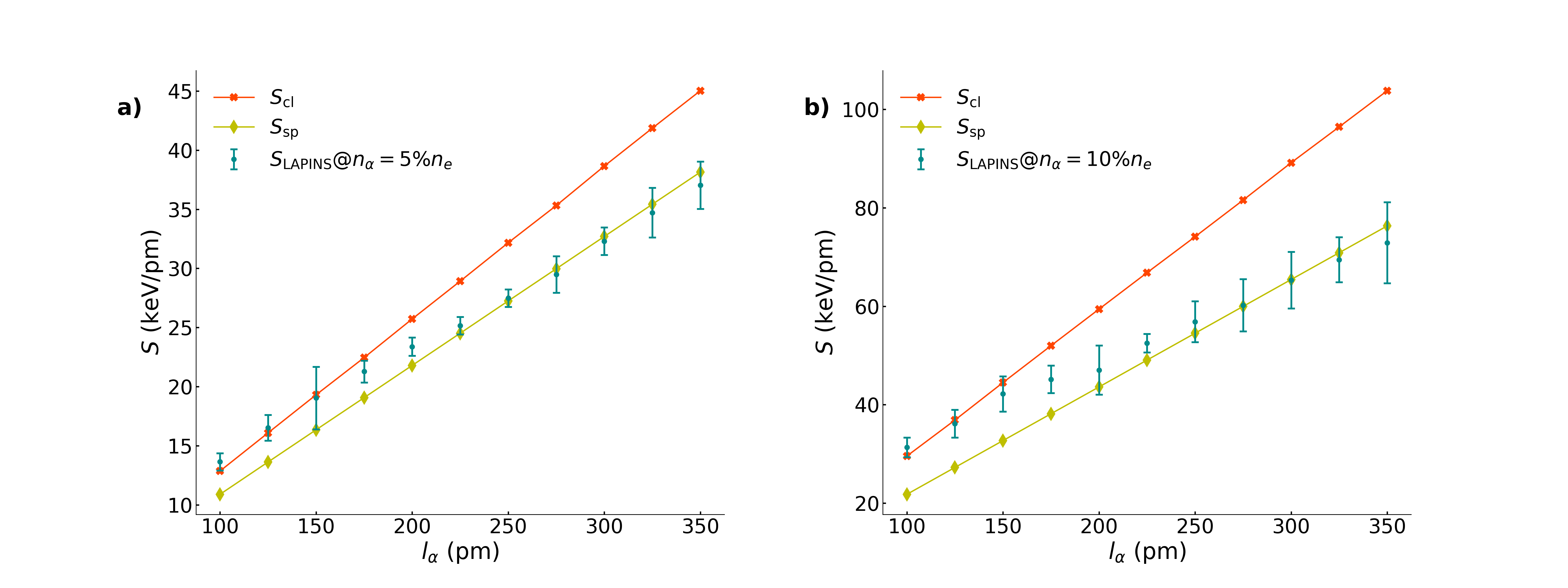}
    \caption{Stopping power of two alpha clusters densities with different sizes, a) \label{fig:0.005ne} simulation results when $n_\alpha=0.005 n_e$, b) \label{fig:0.010ne} simulation results when $n_\alpha=0.01 n_e$. The meaning of lines are the same as Fig.~\ref{fig:FixedSize}.}
    \label{fig:FixedN}
\end{figure*}

\subsection{Numerical methods}
We carried out a series of simulation in verification of our theory.
The simulation is carried out using the PIC code LAPINS \cite{r7,r8,r9}. In LAPINS, multiple physical effects such as collision \cite{r10}, ionization \cite{r11,r12}, radiation \cite{r8}, QED \cite{r13}, and nuclear reactions \cite{r14} are included and coupled. 
The long-range Coulumb interaction and short-range collisions were considered in the simulation. Despite this, in order to take quantum degeneracy into account, Boltzmann-Uehling-Uhlenbeck equation is adopted for the kinetics of electrons, and Fermi-Dirac distributions and the Pauli-exclusion principle among and with electrons are naturally fulfilled. We constrained the distribution of electron energy into a Fermi-Dirac distribution discretized with 20 bins within the range $0\sim2000~\mathrm{eV}$.

In our simulation, we used a 1D simulation box with 240 grids in the longitudinal direction, each grid of length $7~\mathrm{pm}$ with 2000 particles placed such that $\mathrm{D-T}$ ice of density $500~\mathrm{g/cm^3}$, or equivalently plasma of electron density $n_e=1.1958\times10^{32}~\mathrm{m^{-3}}$ and temperature $k_B T_0 = 372.8~\mathrm{eV}$
were simulated effectively. At the initial time an alpha particle beam with max density $n_\alpha$ and cluster longitudinal length $l_\alpha$ is ejected along the longitudinal direction of the simulation box at the energy of $E_k=3.52~\mathrm{MeV}$, a typical value for DT fusion process products. To improve numerical stability, we wrapped the incident cluster with trigonometric-shaped density change of which rising and dropping times are $1.67\times 10^{-3}~\mathrm{fs}$.
The alpha particles were simulated with 400 effective particles per cell.
Absorbing boundary conditions are set at the incidence plane and the end of the simulation box.

\section{Results and Discussion} \label{sec:discuss}

In this section, we will illustrate the applicability of a linear response model based on Coulomb electron gas at finite temperatures, and qualitatively explain the decoherence process of the weak field presented in the results.

\subsection{Stopping power of alpha clusters with different sizes and densities} \label{sec:discussA}

We consider the incident alpha particle as spherical shells with thicknesses $125~\mathrm{pm}$ and $300~\mathrm{pm}$, while their densities are taken from $0.01n_e$ to $0.15n_e$ as a demonstration for the high density of incident particles in the burning plasma, where $n_e$ is the electron density of the DT fuel. 

We used LAPINS simulation results to calculate the stopping power of alpha clusters as shown in Fig.~\ref{fig:FixedSize}, illustrated with blue circles and error bars.
In order to quantitatively investigate whether the interaction effects between alpha particles dominate the total stopping power, we employed the dielectric function theory mentioned in Sec.~\ref{sec:model} to separately calculate the stopping power contributions from the interaction effects and the single-particle term with the same physical parameters of the background plasma as used in simulation.

As shown in Fig.\ \ref{fig:125pm}, when the cluster size is relatively small, we can see that the total stopping power $S_{cl}$ in the dielectric function theory matches the simulation results
very well at lower alpha particle densities. Meanwhile, the interference stopping power $I_{ij}$ occupies a significant portion above a certain density of alpha particles. This shows that the stopping power of alpha cluster cannot be estimated from a mere single-particle perspective. It requires the introduction of effective interactions induced by high-density clusters. However, when the density of alpha particles increases, there is a deviation from the two results and the real stopping power comes closer to the single-particle case.

This phenomenon becomes more obvious when the size of the $\alpha$ cluster is much bigger. As shown in Fig.\ \ref{fig:300pm}\, the real stopping power almost matches $S_{sp}$ in the whole density range. 
These two results show some kind of transition behavior which can be interpreted qualitatively as follows. First, the alpha cluster will heat the electrons as it passes through the background plasma, so the denser the particle cluster, the greater the temperature gradient at both ends of the cluster. The temperature gradient will excite density fluctuations in the particle cluster, in other words, the velocity fluctuations of the particles in the cluster because if the particle velocities were the same, the clusters would have maintained a uniform density distribution.
In the framework of a linear response, the temporal oscillations of the response field $\Phi(\vb{r},t)$ depend on the velocity of the source particle $\vb{v}_p$, shown in Eq.\ \ref{eq:response}.
\begin{equation}
    \Phi(\vb{r},t) = \int\frac{d^3\vb{k}}{(2\pi^3)}e^{i\vb{k}\cdot(\vb{r}-\vb{v}_p t)}\frac{4\pi Ze}{k^2\epsilon(\vb{k},\vb{k}\cdot\vb{v}_p,T)}.
    \label{eq:response}
\end{equation}
Therefore, the coherent superposition of the wake field excited by different particles will be destroyed due to the velocity fluctuations. 

Second, the size of the cluster affects the rigidity of the cluster as a whole and thus affects the amplitude of the velocity fluctuation inside the cluster. As shown in Fig.\ \ref{fig:figure1}\, the force range of the interaction between particles in the cluster is about $0.1~\mathrm{nm}$. Hence, when the size of the cluster is comparable to the force range, the cluster as a whole behaves as a solid under external disturbance, and the density fluctuation and velocity fluctuation in the cluster are usually long-range, so as to better protect the coherent superposition of the wake field. However, as the size of the cluster increases, the cluster as a whole behaves as a fluid under external disturbances, making coherence more vulnerable under disturbances.

\begin{figure}[h]
  \includegraphics[scale=0.375]{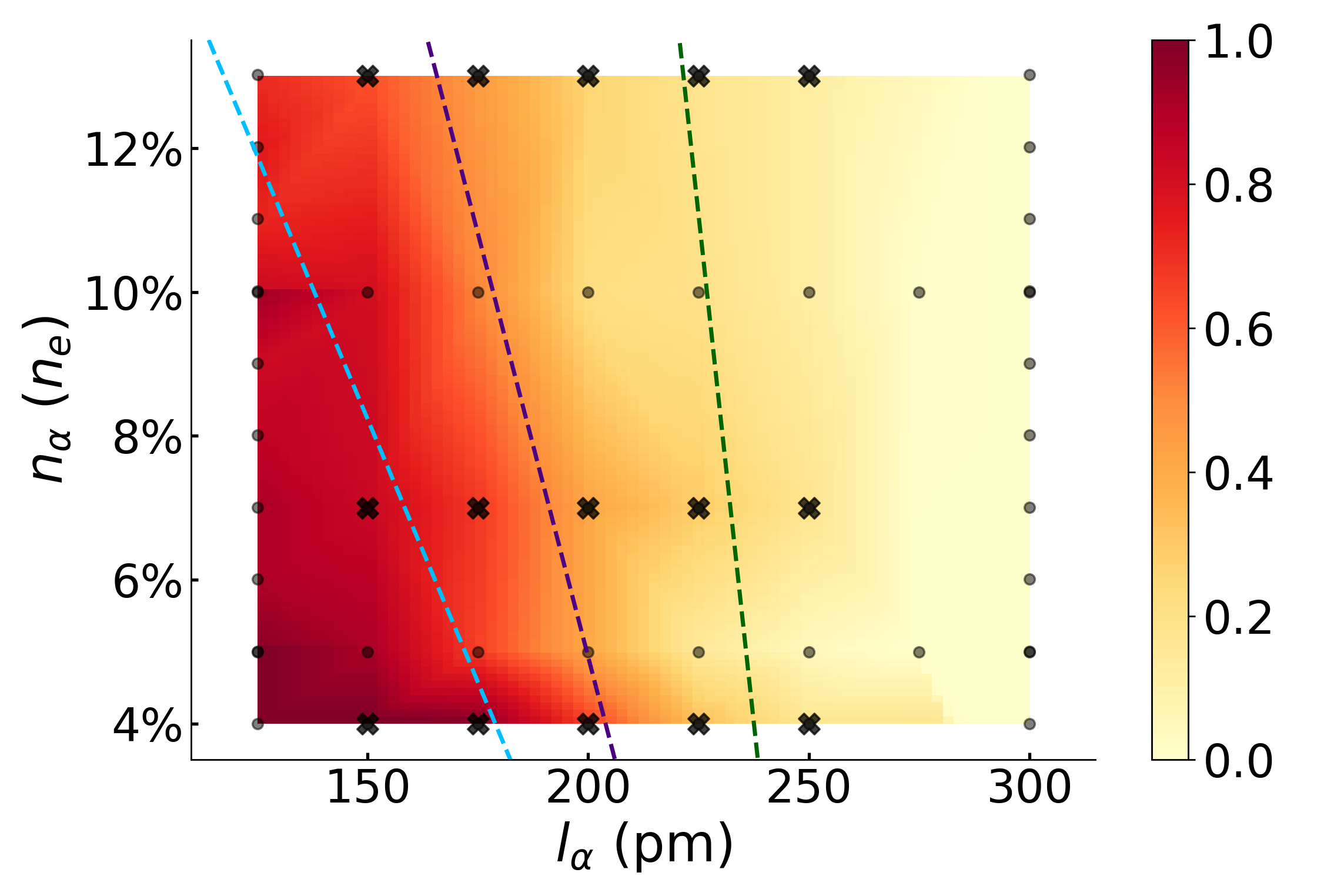}
  \caption{Scope of the dielectric interaction model.
  Black dots and crosses represent parameter points where $\delta$ was drawn directly from simulation, whose results control the linear interpolation of $\delta$. Data from black crosses, in particular, was simulated with 1000 effective particles per grid for electron and plasma ions respectively, and 10 effective particles per grid for $\alpha$ particles. The red-yellow colored value stands for the interpolated result of $\delta$. The light blue dashed line tells the linear regressed borderline where $\delta$ reduces to 80\% with increasing cluster size and density, indicating a 20\% deviation from the dielectric interaction model. Similarly, the dark green dashed line indicates a 20\% deviation from the single particle model. The dark purple line implies the bisector for the scope of the two models.}
  \label{fig:paramspace} 
\end{figure}

\subsection{Scope of the dielectric interaction model}
In order to clarify the decoherence process described in the previous section and determine the parameters under which the effect occurs and ends, we further calculated the stopping power under other parameters. Two of these results are shown in Fig.\ \ref{fig:FixedN}\, illustrating how the stopping power changes with cluster size at a certain cluster density. Also in these two results, we can see a clear transition from coherence to decoherence as the stopping power gradually changes from $S_{cl}$ form to $S_{sp}$ form.

We investigated a rough scope for the interaction model with
$$
\delta = \frac{\abs{S_\mathrm{LAPINS} - S_\mathrm{sp}}}{S_\mathrm{cl}-S_\mathrm{sp}}
$$
by interpolating $\delta$ over simulated points in the parameter space. 
As shown in Fig.\ \ref{fig:paramspace}, our interaction theory gives a significantly better estimation of the stopping power of alpha clusters when the parameters of the cluster are below the line determined by $(l_\alpha,~n_\alpha)=(12\%n_e,~125~\mathrm{pm}),~(4\%n_e,~175~\mathrm{pm})$. Besides, the effect of interaction lasts in the relatively larger cluster and higher density region, making the single particle calculation yet to be suitable for evaluating the actual stopping power, while the dielectric interaction model still serves as a good approximation under the perturbing effect of large temperature gradient induced forces and non-linear effects we mentioned in Sec.\ \ref{sec:discussA}\ . As the cluster size increases to $\sim 250~\mathrm{pm}$, the interaction gradually fades out and the single particle term begins to dominate just as we predicted before.


\section{Summary}\label{sec:sum}
In this paper, we have investigated the stopping power of alpha clusters in WDMs under the typical settings of the ICF process, which illustrates the importance of the effective interaction between particles.
Theoretical analysis and simulation results using LAPINS show that the collective stopping power of the alpha cluster is indeed increased via coherent superposition of excitation fields.(the excitation of high-amplitude wake waves. )
However, the comparison between simulation and theoretical results also illustrate a decoherence transition of the stopping power of the cluster.
The initial condition with different sizes and densities of the alpha clusters have been considered to verify the condition of decoherence transition. At larger cluster size and higher density, we demonstrated that a mere single particle stopping power model would instead give a better estimation due to the coherence breaking induced by velocity fluctuations and the fast decay of stopping force range. Our work can shed light on the ignition and burning process in ICF, point out the conditions for which the interaction theory and single particle evaluation are suitable respec- tively, as well as pose questions on an elaborated theoret- ical model for certain cases with results that fall out of the linear response theory.

{\color{blue}Undergraduate students Z. P. Fu, Z. W. Zhang and K. Lin contributed equally to this work.} Z. P. Fu focused on theoretical analysis, Z. W. Zhang focused on the one-dimensional simulations by the LAPINS code, and K. Lin focused on drafting the manuscript.
\section{ACKNOWLEDGMENTS}
This work is supported by the Strategic Priority Research Program of Chinese Academy of Sciences (Grant Nos. XDA25010100 and XDA250050500), National Natural Science Foundation of China (Grants No. 12075204), and Shanghai Municipal Science and Technology Key Project (No. 22JC1401500).


\end{document}